# Public and private blockchain for decentralized digital building twins and building automation system

Reachsak Ly, Ph.D.[1] Alireza Shojaei, Ph.D.[2]

[1] Assistant Professor, School of Technology, Eastern Illinois University, Charleston, Illinois, United States
[2] Assistant Professor, Myers-Lawson School of Construction, Virginia Polytechnic Institute and State University, Blacksburg, Virginia, United States

**Abstract:**

The communication protocols and data transfer mechanisms employed by IoT devices in smart buildings and corresponding digital twin systems predominantly rely on centralized architectures. Such centralized systems are vulnerable to single points of failure, where a malfunction can disrupt its operational processes. This study introduces a blockchain-based decentralized protocol to enhance the cyber resilience of IoT data transfer for digital twins and enable decentralized automation of building operations. The framework incorporates public and private blockchain technologies alongside two case studies showcasing prototypes of each system. These prototypes were validated within a real-world building environment using smart home appliances and two digital twin platforms, with their performance evaluated based on cost, scalability, data security, and privacy. The findings reveal that the Hyperledger Fabric-based system excels in terms of scalability, speed, and cost-effectiveness while both frameworks offer advantages over traditional centralized protocols in the system's cyber resilience and data security and privacy.

**Keywords**: Blockchain, Digital twin, Building automation system, Hyperledger fabric, Ethereum

1. **Introduction**

A significant portion of people's time is spent in indoor environments, with studies estimating that approximately 90% of human life is spent inside buildings (Latifah et al. 2020). The indoor environments greatly influence occupant's comfort, health, and productivity. Comfortable indoor conditions enhance physical and mental well-being, reducing stress and increasing the occupant's overall satisfaction and work performance (Wyon 2004). Cyber-physical systems (CPS) such as smart buildings play a crucial role in providing a comfortable building environment through indoor environment optimization and efficient building operations, thereby enhancing occupants' satisfaction, productivity, and well-being (Dong et al. 2019). One of the critical aspects of any smart building is its integral connection to smart facility management (FM), which encompasses different tasks, including building automation, building operation and maintenance, energy management, and indoor environmental management. Building automation is enabled by a combination of technologies such as building management systems, sensors, actuators, and IoT devices, among other technologies (Havard et al. 2018). Also, digital building twins, physical building data, and automated building operations within the smart building are all interconnected. Building environmental data can be captured with networks of IoT sensors and used to trigger the automation of specific operations within a building before being stored and later analyzed by other advanced technology (e.g., artificial intelligence ) for further insight (Ye et al. 2018).

However, the current communication protocols and data storage for IoT devices used in smart buildings and their corresponding digital twins (DT) systems are mainly based on the traditional centralized system and database (Lohia et al. 2019; Karaarslan and Babiker 2021).  Such centralized systems are vulnerable to a single point of failure (Agrawal et al. 2018), a vulnerability where the failure of a single component can lead to system-wide disruption, which can be exploited by cyber threats (de las Morenas et al. 2020). Previous studies reveal potential loopholes within traditional building automation systems (BAS), making the system susceptible to data breaches, tempering, forgery, and cyber-attacks such as spoofing, information eavesdropping, and denial of service attacks (Graveto et al. 2022). According to a comprehensive examination of smart building security conducted by Kaspersky in 2019, nearly four in ten (37.8%) automation systems in smart buildings have experienced malicious cyber-attacks (Kaspersky 2021). As per IBM Security report, cyberattacks such as ransomware rose 41% in 2022, with the average cost per data breach being $4.45 million (IBM 2023). Such attacks not only compromise building cyber-physical systems' data integrity but also disrupt its operational processes, leading to unexpected system downtime, compromised functionalities, potential safety risks, discomforts, and privacy concerns for the occupant (Li et al. 2023). Therefore, it is evident that research in enhancing the IoT and digital twin data security and privacy, and the resilience of building automation operations, is crucial in addressing the vulnerabilities and inefficiencies of the current smart building systems.

Blockchain technology provides a potential solution for tackling these challenges within smart building systems. The blockchain serves as a public digital ledger where data is stored and maintained in an immutable and transparent manner within the decentralized peer-to-peer network (Naderi et al. 2024). Blockchain's inherent security features, such as cryptographic security and data immutability can enhance the protection of IoT and digital twin data. Blockchain technology removes the single point of failure within the traditional centralized systems, thereby increasing the system's resilience against cyber threats. Furthermore, blockchain's smart contract is capable of creating decentralized automation and self-executing agreements that can enhance the efficiency and resiliency of building automation operations.

This study seeks to develop a decentralized protocol to improve the resiliency of the transfers of IoT data to the digital twins as well as enable the decentralized automation of building operations. The objectives of this study include: (1) Creating a secure and resilient mechanism for transmitting and storing IoT sensor data and digital twin-related information using both the public and private blockchain network alongside the decentralized oracle network and the Interplanetary File Storage systems (IPFS). (2) Enable decentralized automation of building operations with blockchain smart contracts or chain codes. (3) Create two digital twin systems that are fed by decentralized sensor data streams to facilitate real-time visualization of building environmental conditions. (4) Conduct two case studies for the decentralized digital twin and building automation system using the private blockchain (Hyperledger Fabric) and public blockchain (Ethereum blockchain) (5) Assess the effectiveness and suitability of the proposed blockchain-based systems for the decentralized digital twin and building automation system.

The remainder of the paper is structured as follows. Section 2 provides the literature review on distributed ledger technologies, including the public and private blockchain network, as well as the related works on the blockchain application in the digital twin domain and smart building before outlining the corresponding research gaps. Section 3 describes the research methodology of this study, and Section 4 presents the proposed decentralized framework. The two case studies with corresponding technical implementations are presented in Section 5. Section 6 discusses the findings, evaluates the proposed system, and explores the research's implications and limitations. Section 7 outlines future research directions. Finally, the conclusion is presented in Section 8.

## 2. Background and related works
### 2.1. Blockchain technologies

Blockchain is a digital public ledger where all data is written and stored transparently, securely, and in a tamper-resistant manner within a decentralized network (Hunhevicz and Hall 2020). Transactions are verified and added to the ledger using different consensus mechanisms to guarantee consensus among network participants (Li and Kassem 2021). Blockchain structures information into blocks that are interconnected to form a continuous chain. After data validation, newly created blocks are added to the blockchain in an immutable, time-ordered sequence with the use of hash codes. This architecture makes it difficult to modify the contents of a block, as any changes would render all following blocks invalid. Each block contains a header and body data. The header includes crucial information such as the previous block's hash, timestamp, Merkle root, nonce, and version number, while the body contains the actual transaction data (Yaga et al. 2018).

#### 2.1.1. Private vs Public Blockchain Network

Blockchain networks can be broadly grouped into public and private systems, each with distinct features and use cases. Public blockchains operate as open networks that allow anyone to join, view transactions, and engage in the consensus process (Sai et al. 2021). These networks prioritize decentralization and transparency, offering high levels of security through their extensive node distribution. Public blockchain's consensus mechanisms include Proof-of-Work (PoW) or Proof-of-Stake (PoS) (Sriman et al. 2021), which facilitate the peer consensus on the state of transactions. This approach enhances security but can lead to slower transaction processing rates and scalability challenges (Lashkari and Musilek 2021).

Private blockchains, which are also referred to as permissioned networks, limit access to authorized participants. These systems offer greater control over network governance, faster transaction processing, and enhanced privacy. Private blockchains can achieve consensus more rapidly due to their limited number of trusted participants, allowing for higher transaction throughput (Al-Sumaidaee et al. 2023). They also provide stronger data privacy controls, as changes can be implemented more flexibly when all authorized nodes agree. Private blockchains are particularly suitable for enterprise applications where data confidentiality and regulatory compliance are important. While public blockchains excel in scenarios requiring maximum decentralization and censorship resistance, private blockchains provide benefits such as scalability, privacy, and customization for specific business needs (Yang et al. 2020).

### 2.1.2. Ethereum blockchain

The introduction of Ethereum in 2015 marked a significant milestone in public blockchain. Unlike its predecessor, Bitcoin, which primarily focused on cryptocurrency transactions, Ethereum expands blockchain functionality to become a versatile platform designed for running smart contracts and creating decentralized applications (DApps) (Tas and Tanriover 2019). This innovation has substantially broadened the possibility of blockchain applications, facilitating the decentralization of markets and enabling complex, programmable transactions. The Ethereum network operates on a proof-of-stake (PoS) consensus system, which delivers enhanced energy efficiency and scalability relative to proof-of-work (PoW) (Zhang and Anand 2022). In this mechanism, the validating peers are selected to generate new blocks by staking their cryptocurrency as collateral (Nguyen et al. 2019).

Ethereum's architecture is built around two account categories: externally owned accounts (EOAs) and smart contract accounts (Li et al. 2020). EOAs are controlled by users, which can be controlled by private keys and can hold Ether, the native cryptocurrency of the Ethereum network. These accounts can execute transactions with other EOAs or engage with smart contracts. Contract accounts, on the other hand, are controlled by their code and are executed when they receive messages from other accounts. In addition, a defining feature of the Ethereum blockchain is the smart contract, a self-executing code with established rules embedded within the blockchain (Naderi et al. 2023). Smart contract's functions can be designed to be self-executed when specific conditions are met, which enables complex, trustless interactions without intermediaries. Smart contracts are developed in the Solidity programming language before being compiled into bytecode and executed using the Ethereum Virtual Machine (EVM) (Oliva et al. 2020).

### 2.1.3. Hyperledger fabric blockchain

Hyperledger Fabric is a private blockchain network under the family of the Hyperledger project that was established by the Linux Foundation in 2016 (Androulaki et al. 2018). Hyperledger Fabric's modular architecture offers a unique approach to private, permissioned blockchain networks, making it particularly suitable for business applications that require fine-grained access control with improved flexibility and scalability (Nasir et al. 2018).

The platform's architecture is composed of several key concepts and components.

- Identity: Unlike public blockchains, every Hyperledger Fabric node has a unique identity, which is categorized into three main roles: (i) Clients: Responsible for submitting transaction proposals and sending transactions for ordering. (ii) Peers: Transaction execution and validation. (iii) Orderers: Aggregate transactions from clients and establish their final transaction sequence (Manevich et al. 2018).
- Membership Service Provider: Responsible for overseeing identities and access permissions within the network, ensuring that only authorized peers can engage with the blockchain system. The Membership Service Provider works in conjunction with the Fabric CA to manage identities and define the rules for their validation (Manevich et al. 2019).
- Transaction: Transactions in Hyperledger Fabric follow a three-phase process: proposal, endorsement, and commitment. A client application first submits a transaction proposal, which is endorsed by a subset of peers. The endorsed proposal is subsequently ordered into a block by the ordering service and finalized in the ledger by all peers. This process ensures that all transactions are validated and recorded consistently.
- Chaincode: Chaincode contains the business logic that governs transactions on the blockchain. Chaincode is deployed on peer nodes and can be invoked by client applications to read from or write to the ledger (Foschini et al. 2020).
- Channels: Fabric provides a mechanism for creating separate communication layers within the network through channels. Each channel has its own ledger and set of chaincodes, and participation in a channel is governed by the Membership Service Provider. This allows for the creation of private subnets of communication between specific network participants, enhancing privacy and scalability (Surjandari et al. 2021).
- Ledger: The ledger is made up of two elements: the blockchain and the world state. The blockchain is an append-only log of all transactions, whereas the world state functions as a database that stores the current state of the ledger. Each peer node maintains a copy of the ledger so that all transactions can be verified by any participant in the network (Nasir et al. 2018).

### 2.2. Digital building twins

A digital twin is fundamentally a virtual replica of a physical asset, system, or process. It serves as a dynamic reflection of its physical counterpart, which is continuously updated through data collected by sensors and IoT devices (Jones et al. 2020). The digital twin concept comprises three essential elements: the physical components, its virtual counterpart, and the bi-directional data flow between them. Digital building twins are virtual representations of physical buildings that integrate real-time data to simulate and monitor the building's performance, condition, and behavior (Hunhevicz et al. 2022). It integrates BIM models, the Internet of Things, and data streams to create a live digital representation of a physical structure by providing a live view of a building's performance, condition, and usage patterns. This real-time data integration can facilitate predictive maintenance, energy optimization, and a decision-supporting system within the building's lifecycle. Specifically, in the building operational phases, digital building twin is essential in enabling data-driven facilities management, thereby improving building operational efficiency and sustainability and enhancing occupant's living experience (Zhao et al. 2022).

### 2.3. Blockchain and the Internet of Things

Over the years, several research works have focused on leveraging blockchain technology to address the vulnerability of IoT communication and data security. Previous research has demonstrated the use of the Ethereum blockchain in developing distributed access control for IoT devices to reduce the centralized nature of access management (Novo 2018). For instance, Fakhri and Mutijarsa (2018) implemented a public Ethereum blockchain-based communication protocol for the IoT system and conducted a comparative analysis with the traditional MQTT-based IoT communication method. They perform security testing on both systems using the simulated attack and sniffing attacks by utilizing the Wireshark software. The findings indicate that the IoT system incorporating blockchain technology is more cyber-resistant compared to the traditional MQTT-based system. Hasan et al. (2022) created an IoT data streaming application with the Ethereum blockchain and the IPFS. To demonstrate the security of the system and resilience to cyber threats and vulnerabilities, the Oyente tool was used to perform security analysis on the smart contract code.

However, one of the main technical limitations of the above research is the scalability of the Ethereum blockchain which includes transaction latency and low throughput (Bez et al. 2019). These constraints may reduce the inefficiency and robustness of the IoT sensor data transfer, especially in the context of real-time data transmission. To address this problem, researchers have leveraged different types of private blockchain networks for IoT data transfers. Research by Balakumar and Kavitha (2021) utilized the Quorum blockchain with the IPFS to secure the data privacy and security of IoT devices. In addition, Iftekhar et al. (2021) demonstrated the feasibility of using Hyperledger Fabric for IoT sensor data transmission. Their work involved implementing attribute-based access control and adapting the Hyperledger Fabric for ARM64 architecture by using Raspberry Pi.

### 2.4. Blockchain technology for digital twin and smart building operations.

Different studies have integrated blockchain technology with digital twins and IoT within the context of building operations and facility management (Adu-Amankwa et al. 2023). For instance, research by Hunhevicz et al. (2022) leveraged digital twins and smart contracts in creating performance-based payment methods in smart buildings with real-time building performance data. The technical implementation of the framework includes smart contracts, digital building models, IoT sensors, digital twin platforms, and Ethereum blockchain. The research highlights the potential of leveraging crypto-economic incentives and performance-based smart contracts to create a peer-to-peer economy in the built environments. In another study, Pittaras et al. (2024) leverage the Web of Things standards and blockchain technology to develop digital twins for IoT devices. The authors present two variations of their digital twin design using both the Ethereum and Hyperledger fabric blockchain before discussing the associated trade-offs. This dual approach allows for flexibility in application depending on specific use-case requirements. In addition, EtherTwin (Putz et al. 2021) aims to offer the secure storage of digital twin data using the Ethereum blockchain. The proposed system allows participants from different building lifecycles to create, modify, upload, and store digital twin-related documents in a decentralized way. Recent studies have also integrated blockchain technology and digital twins in smart building environments. For instance, Teisserenc and Sepasgozar (2022) emphasize that blockchain-powered digital twins can strengthen the decentralization of building data by enhancing data traceability, security, and privacy. Integrating blockchain with digital twins could empower facility managers to conduct real-time asset monitoring and contribute to more efficient automation and prediction of maintenance activities.

In addition, a recent study by Jeoung et al. (2022) investigated the creation of a blockchain-integrated IoT framework designed to enable personalized indoor temperature regulation within building management systems, with a focus on ensuring data privacy and security. Furthermore, Tiwari and Batra (2021) leveraged smart contracts to enhance maintenance requests and facility management tasks within a building infrastructure. This blockchain-based system can minimize dependence on third-party management services and enhance automation and transparency in building infrastructure. A study by Xu et al. (2018) presents a proof of concept of the decentralized smart home system utilizing Raspberry Pi, the Blynk platform, DHT11 sensors, and the Ethereum blockchain. The live environmental data such as temperature and humidity data will be recorded on the blockchain network and checked against the threshold before triggering a certain action (e.g., LED warning). Also, Majeed et al. (2020) and Umer et al. (2023)

use the Ethereum blockchain for the identity management application of IoT devices within the smart home system. Additionally, a study by Ly et al. (2024) proposed the concept of a data-driven and decentralized governance system in smart building facilities management using digital building twins and decentralized autonomous organizations. Their subsequent study further advanced this concept with the integration of a large language model and blockchain-based governance for smart building automation (Ly and Shojaei 2024).

These studies emphasize the potential of blockchain and digital twin technologies to revolutionize facility management practices, offering promising avenues for improving building operational efficiency and data security in smart building environments. The works have recognized the significance of IoT and digital twin data security and the potential benefits of blockchain application in smart building facility management and building automation. However, the research on using private blockchains, such as Hyperledger Fabric, in smart building automation systems is still unexplored. Besides, the existing research on integrating decentralized automation of building operation with secured decentralized digital building twin into one integrated system within the building infrastructure.

## 3. Research Methodology

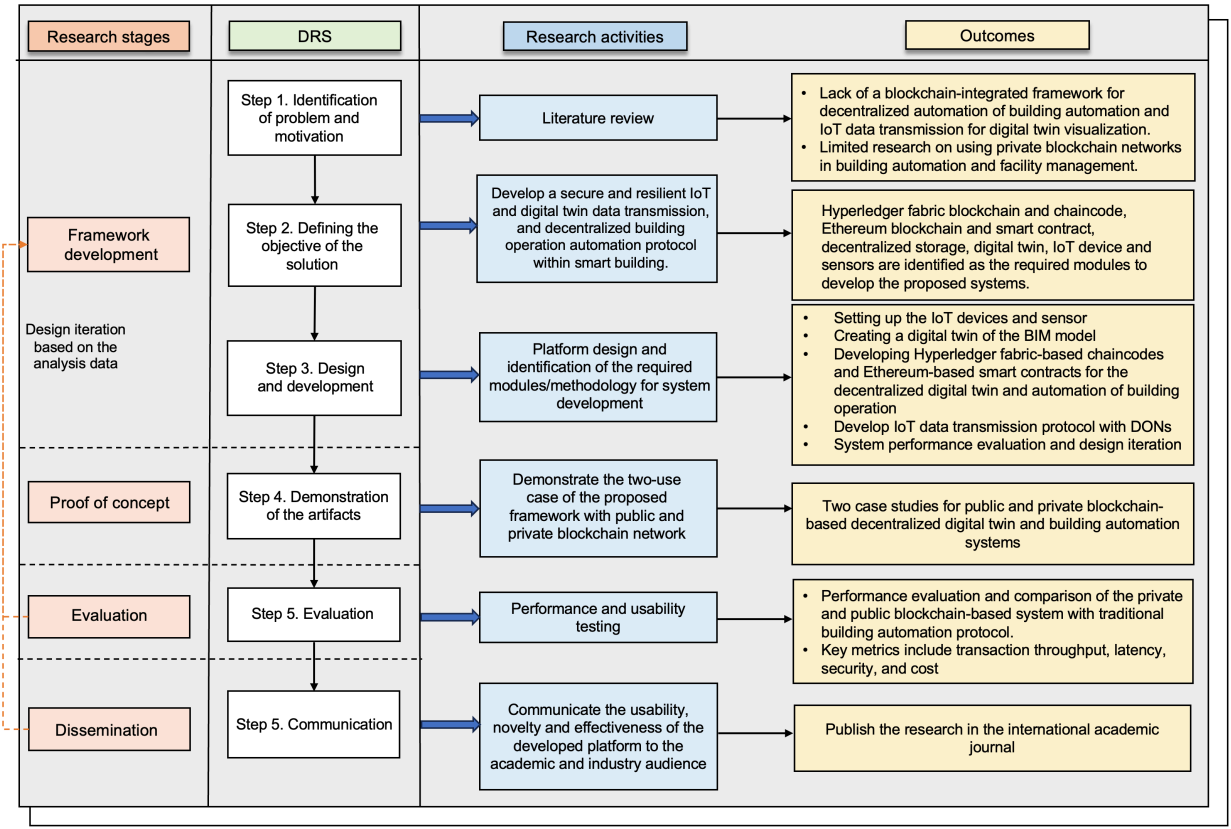

**Fig. 1** DSR-based research flow

This study adopts the Design Science Research (DSR) method, a scientific problem-solving paradigm that involves a systematic process of identifying real-world problems, designing and creating novel artifacts (such as frameworks, methods, or instantiations), evaluating the designed solutions, and contributing to the theoretical body of knowledge (Hevner et al. 2004). Many scholars have adopted the DSR method to develop blockchain-based applications in the construction domain. For instance, Elghaish et al. (2023) used this DSR approach for the development and validation of a blockchain application in construction supply chains. In addition, Wu et al. (2022) utilized DSR to examine the application of blockchain for improving information-sharing efficiency for modular construction projects, while Cheng et al. (2023) utilized DSR to create a blockchain application for construction cost management. The process of DSR comprises six different stages (Geerts 2011). Fig. 1 describes the DSR-based research methodology of this study with different research stages and corresponding DSR steps.

(1) Problem identification and motivation. The literature review in Section 2 has revealed that there is a lack of an integrative framework that can facilitate autonomous and distributed building facility automation, secure real-time IoT data transmission and storage, and digital building twin visualization into a single system within the building infrastructure. Moreover, the research on private blockchain network applications, such as Hyperledger Fabric, in smart building automation and facility management is also limited.

(2) Objective definition. To address this research gap, this study seeks to develop a comprehensive decentralized framework to secure IoT data transfer for decentralized digital twins as well as enhance the resilience of building operations automation by integrating smart contracts, IoT, digital twin, and blockchain technologies.

(3) Design and development. The proposed framework will be designed and developed through the following stages. (i) Setting up the IoT sensors, single board computers, and smart home appliances for data collection and building automation (ii) Creating a digital twin of the modeled building infrastructure (iii) Developing Hyperledger fabric-based chaincodes and Ethereum's smart contract for the IoT data transmission between the IoT edge device to digital twin platform and enable the automation of building facilities. (iv) Leverage the decentralized oracle network and the developed Ethereum-based smart contract for IoT data collection and transmission to the digital twin, as well as automate the building facilities operation. (v) System performance evaluation and design iteration.

(4) Demonstration. Develop the prototypes of the framework using Hyperledger Fabric and Ethereum blockchain through two case studies.

(5) Evaluation. Performance evaluation and comparison of the two blockchain networks implemented in the case studies will be conducted. Key metrics, including transaction throughput, latency, and cost, will be analyzed to determine the suitability and trade-offs of each blockchain platform for the proposed framework. Additionally, a qualitative analysis will be conducted by discussing and comparing key characteristics such as data security and privacy between the traditional building automation protocol such as BACnet with the proposed blockchain systems.

(6) Communication. The design, development, and evaluation results of the proposed system will be published in an academic journal.

## 4. Proposed decentralized framework for building operation automation and data transmission.
### 4.1. Framework Overview

The aim of this study is to develop a decentralized protocol for IoT data transmission for digital twins and enable decentralized automation of building operations within smart buildings using blockchain technology. Fig. 2 provides the overview of the general architecture of the proposed system with various modules working together. The proposed framework encompasses different modules in both the cyber and physical components, with the cyber components consisting of the blockchain network, decentralized storage, smart contracts/chain codes, real-time data visualization, and digital twin components. The physical components encompass physical assets, such as IoT sensors, Raspberry Pi, and building operations. The primary component of the framework is the blockchain network, which can be implemented using either Hyperledger Fabric or Ethereum.

During operation, real-time environmental data will be collected from the building environment through IoT sensors and Raspberry Pi. This data is securely transferred to the blockchain network for further processing and storage. The decentralized storage component, such as the IPFS, will be used for storing the collected historical data. The smart contract/chain code module, implemented on the blockchain network, serves as the decision-making engine for automating building operations. Based on predefined logic and conditions programmed within the smart contracts, specific events and actions will be triggered in response to the incoming sensor data streams. This module communicates with HVAC and lighting systems, to implement the necessary response in the physical world. These actions can include alerts, adjustments to the HVAC system, lighting system, and other critical building operations. In addition, the digital twin platform provides real-time data visualization of the building's environmental condition. This visualization is powered by the live sensor data streams retrieved from the blockchain network.

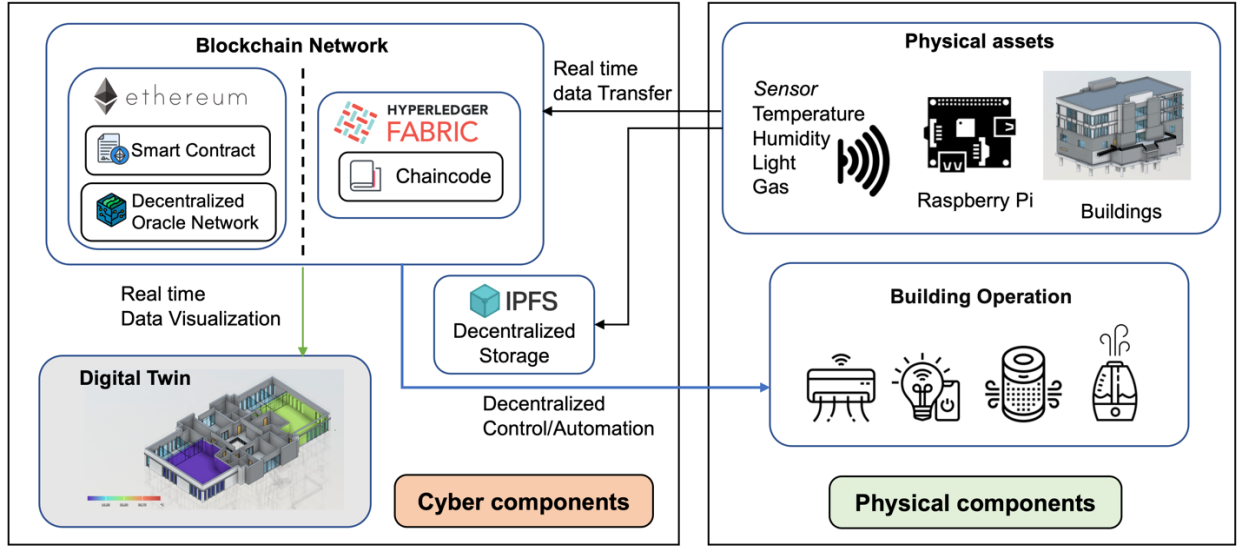

**Fig. 2** Overview of the proposed framework

### 4.2. Blockchain and digital twin components

The proposed framework leverages blockchain technology to enable decentralized data transmission, storage, and automation of building operations. The blockchain components within the Hyperledger fabric-based system in the first case study mainly include the Hyperledger chaincodes. The Ethereum smart contracts and decentralized oracle network are the main components in the Ethereum-based system in the second case study. These components in each blockchain system work together to ensure data integrity, transparency, and efficient automation of building operations. In addition, the digital building twin is composed of both static components (BIM model) and dynamic components (Real-time sensor data streams). The building's digital twin provides a comprehensive digital representation with real-time updates on the building's environmental conditions.

#### 4.2.1. Hyperledger Fabric-based decentralized digital twin and building automation system framework.

In the Hyperledger Fabric-based system, two different chaincodes were developed for smart device automation and sensor data handling within the blockchain network. The first chaincode is responsible for automating facility management tasks based on the sensor data received from the building environment. This chaincode is programmed with predefined rules and logic to process the incoming sensor data streams. It retrieves the sensor data, such as temperature, humidity, light intensity, and carbon monoxide concentration, before comparing them against predefined thresholds. Based on these comparisons, the chaincode triggers appropriate actions, such as activating smart fans, heaters, lights, air purifiers, or humidifiers. The second chaincode facilitates the secure sensor data transfer from the blockchain to the digital twin platform to enable decentralized real-time visualization of building environmental conditions. Additionally, all historical sensor data is stored on the IPFS to ensure a decentralized, secure, and immutable archive of historical building environmental data. In addition, Autodesk Tandem is chosen for the digital twin implementation for this system. Autodesk Tandem is an out-of-the-box software solution for digital twin applications that requires minimal programming skills (Autodesk 2023). It provides a comprehensive platform for creating and managing digital twins of physical buildings by providing seamless integration of BIM data and IoT sensor streams. The overview of the proposed Hyperledger Fabric-based system is illustrated in Fig. 3.

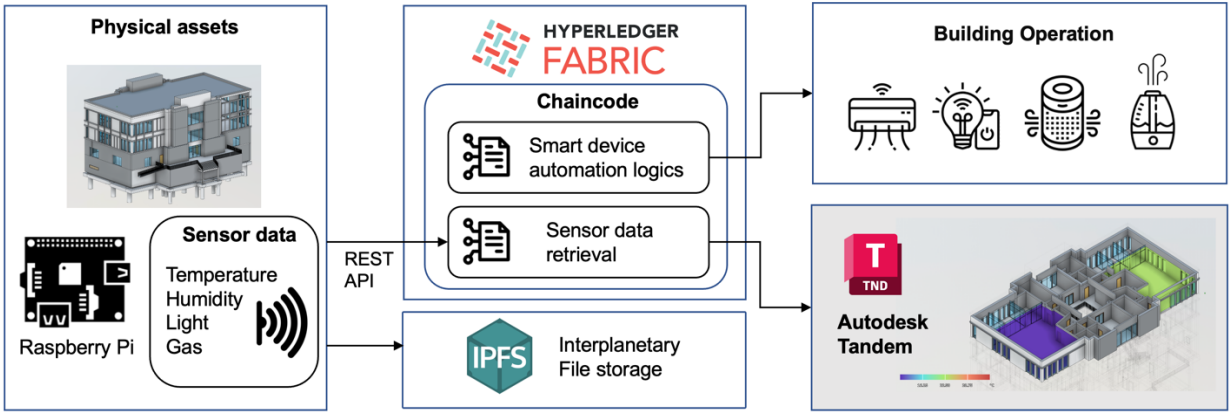

**Fig. 3** Overview of the proposed Hyperledger Fabric-based system for decentralized digital twin and building automation system

### 4.2.2. Ethereum-based decentralized digital twin and building automation system framework

In this Ethereum-based system, smart contracts are the core components of the automation of smart building operations and the creation of the decentralized digital twin. The use of Ethereum smart contracts ensures that these real-time data, building automation-related threshold parameters, and operations are decentralized, thereby enhancing the security and resilience of the automation process.

There are two main smart contracts in this framework. The first smart contract is designed to store real-time environmental data from the building environment before updating the digital building twin. The second smart contract is designed to store the predefined threshold parameters, which will be used to compare against the real-time environment data for the building automation logic. This contract enables decentralized and automated control over the smart devices based on real-time sensor data inputs.

In addition, this study opted for Autodesk Platform Services instead of Autodesk Tandem for the digital twin implementation within this Ethereum-based system. This option allows for greater customization and flexibility in developing the digital twins. In addition, in this proposed framework, the Chainlink decentralized oracle network is used to automatically fetch and verify environment sensor data from the REST API created by the data collection component at specific time intervals. Decentralized oracles facilitate the decentralized data flow from the off-chain data source to the on-chain smart contracts. The use of decentralized oracles ensures that the data fed into the Ethereum smart contracts is tamper-proof and sourced from reliable sources. Fig. 4 illustrates the overview of the proposed Ethereum-based system.

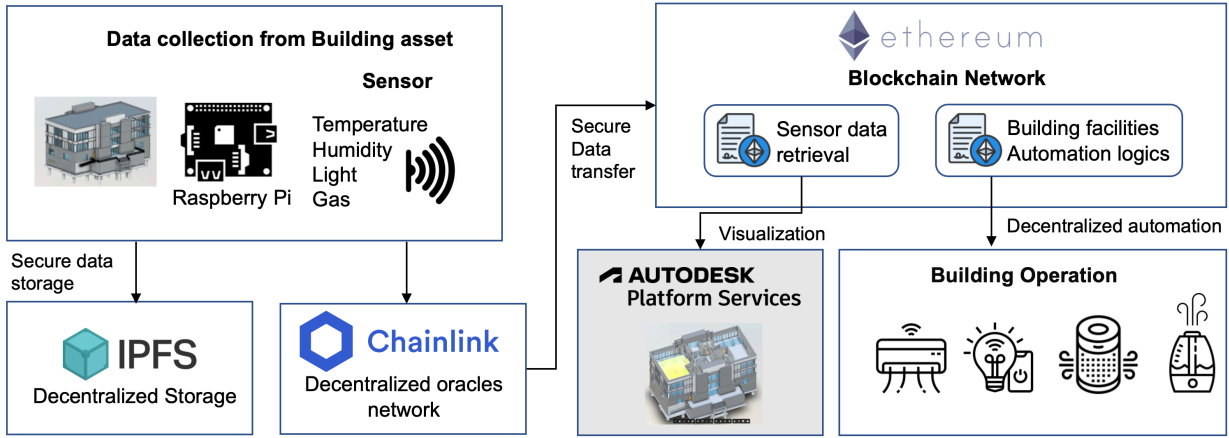

**Fig. 4** Overview of the proposed Ethereum-based system for decentralized digital twin and building automation system

### 4.3. Decentralized storage

To handle the large volumes of sensor data at a lower computational cost, this study utilizes the IPFS as decentralized storage. IPFS is used to securely store and retrieve historical sensor data and build operational data in both frameworks.

The sensor data collected from the building environment is automatically uploaded to the IPFS network at specific time intervals.

### 4.4. Smart building facilities automation

The proposed framework leverages the integration of IoT sensors, smart building facilities, and smart contracts/chaincodes to create decentralized automation of building operations. By continuously monitoring the building's environmental data, the system can intelligently trigger appropriate actions to optimize occupant comfort, energy efficiency, and indoor air quality. These smart contracts or chaincode act as decentralized and tamper-proof controllers, executing specific actions in response to the incoming sensor data streams. The activation of smart building facilities is facilitated through these blockchain-based smart contracts, ensuring a secure, transparent, and automated process. The smart devices incorporated in this system include a smart fan, smart light bulb, smart air purifier, and smart humidifier. The automation logic for these devices is governed by predefined thresholds and facilitated through blockchain smart contracts, ensuring secure and transparent operations.

The automation logic of the smart building system in the proposed frameworks is as follows. (1) Smart HVAC System (Smart Fan and Smart Heater): When the temperature sensor (DHT11) detects a value above a predefined upper threshold, the smart contract triggers the activation of the smart fan to cool the environment. Conversely, if the temperature falls below a lower threshold, the smart contract activates the smart heater to maintain a comfortable indoor temperature. (2) Smart Lighting System (Smart light bulb): The light intensity sensor continuously monitors the ambient light levels within the building spaces. If the detected light intensity level falls below a minimum threshold, the smart contract automatically turns on the smart light bulb to ensure adequate illumination. Conversely, when the light intensity level exceeds a maximum threshold, indicating sufficient natural lighting, the smart contract dims or turns off the smart light bulb to conserve energy. (3) Indoor Air Quality Management (Smart Air Purifier and Smart Air Humidifier): The MQ-2 gas sensor detects the presence of harmful gases, smoke, or air pollutants within the indoor environment. If the gas concentration exceeds a predefined threshold, the smart contract activates the smart air purifier to filter and purify the indoor air. Additionally, the DHT11 sensor monitors the relative humidity levels. When the humidity falls below a minimum threshold, the smart contract triggers the smart air humidifier to maintain optimal humidity levels for occupant comfort.

### 5. Proof of Concept

Two case studies are presented to showcase the implementation and validation of the proposed decentralized framework for building operation automation and IoT data transmission for digital twins. The first case study focuses on a private blockchain-based prototype that operates on the Hyperledger Fabric blockchain network. The second case study examines a public blockchain-based prototype with the use of an Ethereum blockchain network. The technical implementation of the prototypes for both case studies is conducted using various tools, programming languages, and development environments. Table 1 summarizes the tools and technologies employed for each module in the respective prototypes. The code for the technical implementation is publicly available under an open-source license (Ly 2025).

**Table 1** Tools used for prototype development

| Case study | Tasks | Programming language (packages) | Development environment |
|---|---|---|---|
| Hyperledger Fabric-based prototype | Configuration of IoT sensors and smart home appliance | Python | Visual Studio Code |
| | Digital twin development | JavaScript (Autodesk Tandem API) | Visual Studio Code |
| | Chaincode development | JavaScript | Visual Studio Code |
| Ethereum blockchain-based prototype | Configuration of IoT sensors and smart home appliance | Python | Visual Studio Code |
| | Digital twin development | JavaScript (Autodesk Platform service) | Visual Studio Code |
| | Smart contract development | Solidity | Remix IDE |

## 5.1. Implementation Preparation
### 5.1.1. Environmental data collection

To capture the dynamic environmental data required for the digital twin, a network of IoT sensors and devices is deployed within the building environment. In this study, a single-board computer, Raspberry Pi 4B, and connected environmental sensors are used for data collection and processing. (Fig. 5). DHT11 sensor measures the ambient temperature and relative humidity levels within the building environment. Light Intensity sensor detects the level of ambient light within the building spaces. The MQ-2 gas sensor is used to detect the existence of different gases, including carbon monoxide, combustible gases, smoke, and other air pollutants. The data collected will be read and processed on the Raspberry Pi 4B with different Python libraries including Adafruit_DHT (for the DHT11 sensor) and Rpi.GPIO, and Adafruit_MQTT. The processed data will be transmitted to the corresponding blockchain networks via a REST API using Python's Flask library.

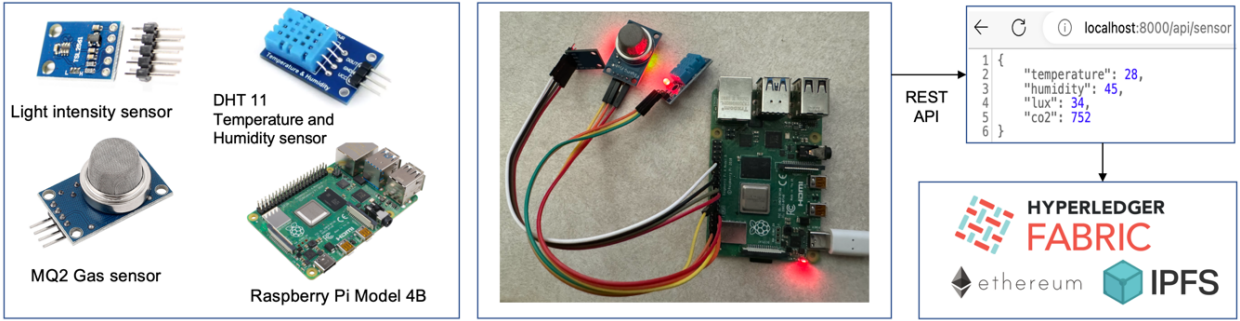

**Fig. 5** Environmental data collection setup

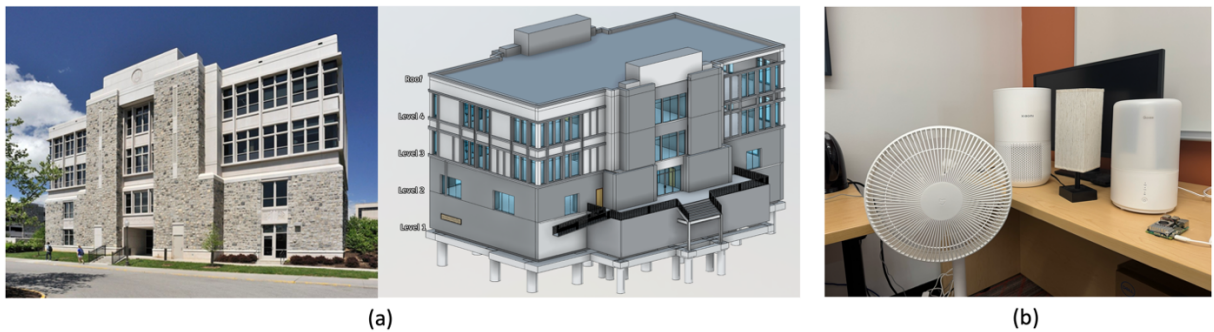

**Fig. 6** a) Virginia Tech Bishop-Favrao Hall and its BIM model b) Smart home appliance

**Fig. 7** a) Chaincode for digital twin b) Bash script for data query c) Bash Script for data synchronization with the digital twin platform

### 5.1.2. Digital building twin development and related equipment

This study selects Bishop-Favrao Hall, home to the Department of Building Construction at Virginia Tech, as the case study location. The BIM model of Bishop-Favrao Hall was developed using Autodesk Revit 2024 (Fig. 6 (a)). To showcase different approaches to digital twin development, two different digital twin platforms were utilized for the case studies. For the Hyperledger Fabric-based digital twin, Autodesk Tandem was chosen due to its minimal coding expertise requirement and provides a ready-to-use solution for integrating building BIM data and IoT data streams. In contrast, for the Ethereum-based digital twin, Autodesk Platform Services was used. This platform offers greater customization and flexibility, allowing for the tailored development of digital twins to meet specific project requirements.

Although the room is equipped with a built-in HVAC system and integrated lighting, these systems are not publicly accessible for direct control. To simulate the proposed system's building operation automation capabilities needed for the study, a variety of smart home devices were installed (Fig. 6 (b)). These devices enable access control for air quality, humidity, lighting, and fan speed through the blockchain-based decentralized system. For air quality management, the Xiaomi Smart Air Purifier 4 Compact was used. It features multiple fan speed options to adjust airflow and purification levels. The Govee Smart Humidifier H7141 was used for humidity level control, as it provides precise adjustments to the room's humidity. The Xiaomi Mi Smart Standing Fan 2 was used to regulate air circulation within the room with customizable fan speed settings that mimic the functions of traditional HVAC systems. Lighting control was achieved using Yeelight Smart Light Bulbs W3, which provide adjustable brightness to replicate indoor lighting management.

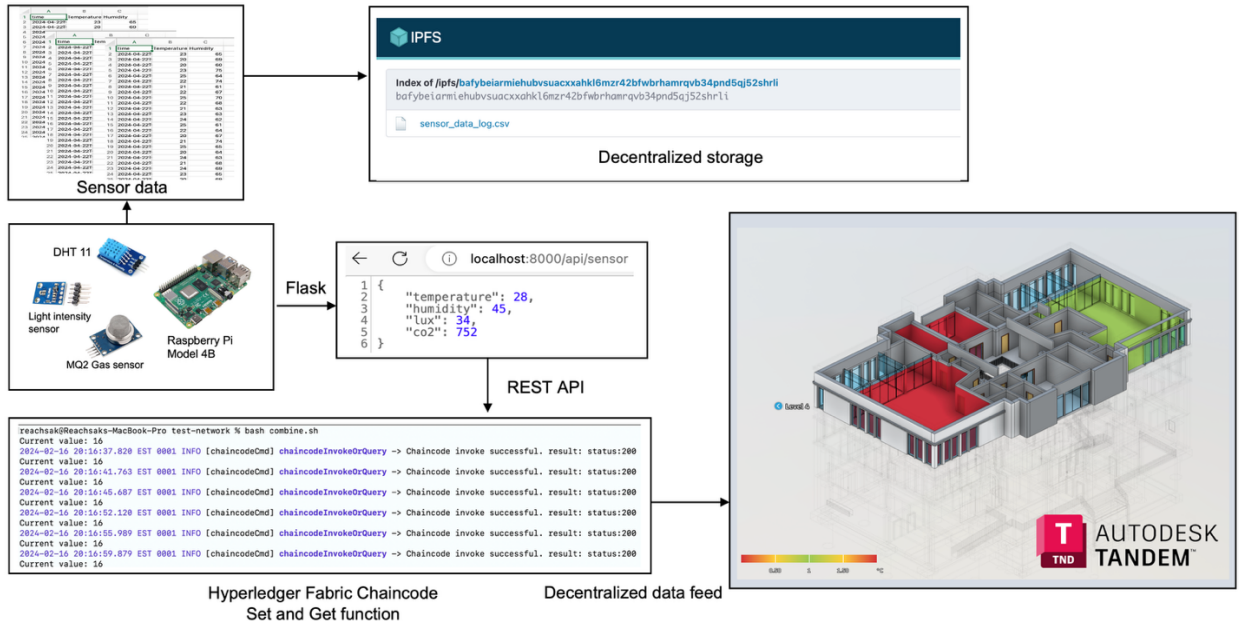

**Fig. 8** Overview of the Hyperledger Fabric-based system for decentralized digital twin

### 5.2. Case study 1
#### 5.2.1. Hyperledger fabric-based digital twin

The Hyperledger Fabric-based digital twin prototype demonstrates the secure and decentralized transmission of sensor data to a digital twin platform in creating a visualization of building environmental conditions. The Hyperledger Fabric's chaincode was developed using JavaScript. The Hyperledger Fabric environment for this prototype is configured to include two peer organizations, namely org1.example.com and org2.example.com. Each organization hosts one peer, with peer0.org1.example.com running at localhost:7051 and peer0.org2.example.com operating at localhost:9051. Communication between these peers is established through a single channel, named mychannel. This simplified configuration is intentionally designed to demonstrate the functionality of Hyperledger Fabric in a straightforward and controlled environment. This setup minimizes complexity while maintaining the core features of a permissioned blockchain network. The deployment process begins with the installation of the chaincode on each peer. The chaincode is instantiated and committed to the channel using Hyperledger Fabric's CLI tools. The deployed chaincode contains two key functions: SetBuildingData and GetBuildingData (Fig. 7 (a)). These functions facilitate the recording and retrieval of building data, such as temperature, humidity, carbon monoxide level, and light intensity level, into and from the blockchain ledger.

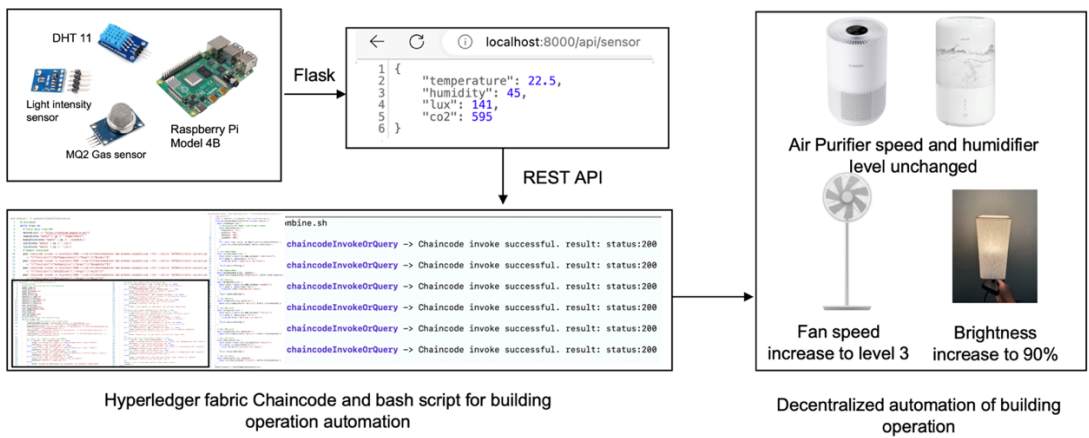

**Fig. 9** a) Chaincode for decentralized building operation automation b) Bash script for data query c) Bash Script for building operation logics d) Hyperledger Explorer

In addition, two bash scripts, as illustrated in Fig. 7 (b) and Fig. 7 (c), are developed to periodically update the chaincode with up-to-date building environment data and synchronize the data from the blockchain chaincode with the Autodesk Tandem API for real-time visualization within the digital twin platform. The first script interacts with the blockchain by invoking the GetBuildingData function, which retrieves a JSON object containing essential building parameters such as temperature, humidity, carbon monoxide level, and light intensity level from the REST API created by the Raspberry Pi. Once the data is retrieved, it is formatted and transmitted to the Autodesk Tandem API through a POST request using the second script (Fig. 8). To maintain up-to-date visualization, the script is designed to operate continuously in an infinite loop, querying and synchronizing data at regular intervals of 60 seconds. In addition, historical sensor data is periodically stored in the IPFS. This approach mitigates the risks of data tampering and leakage. The stored historical data can also be retrieved for later analysis, such as trend analysis and predictive maintenance.

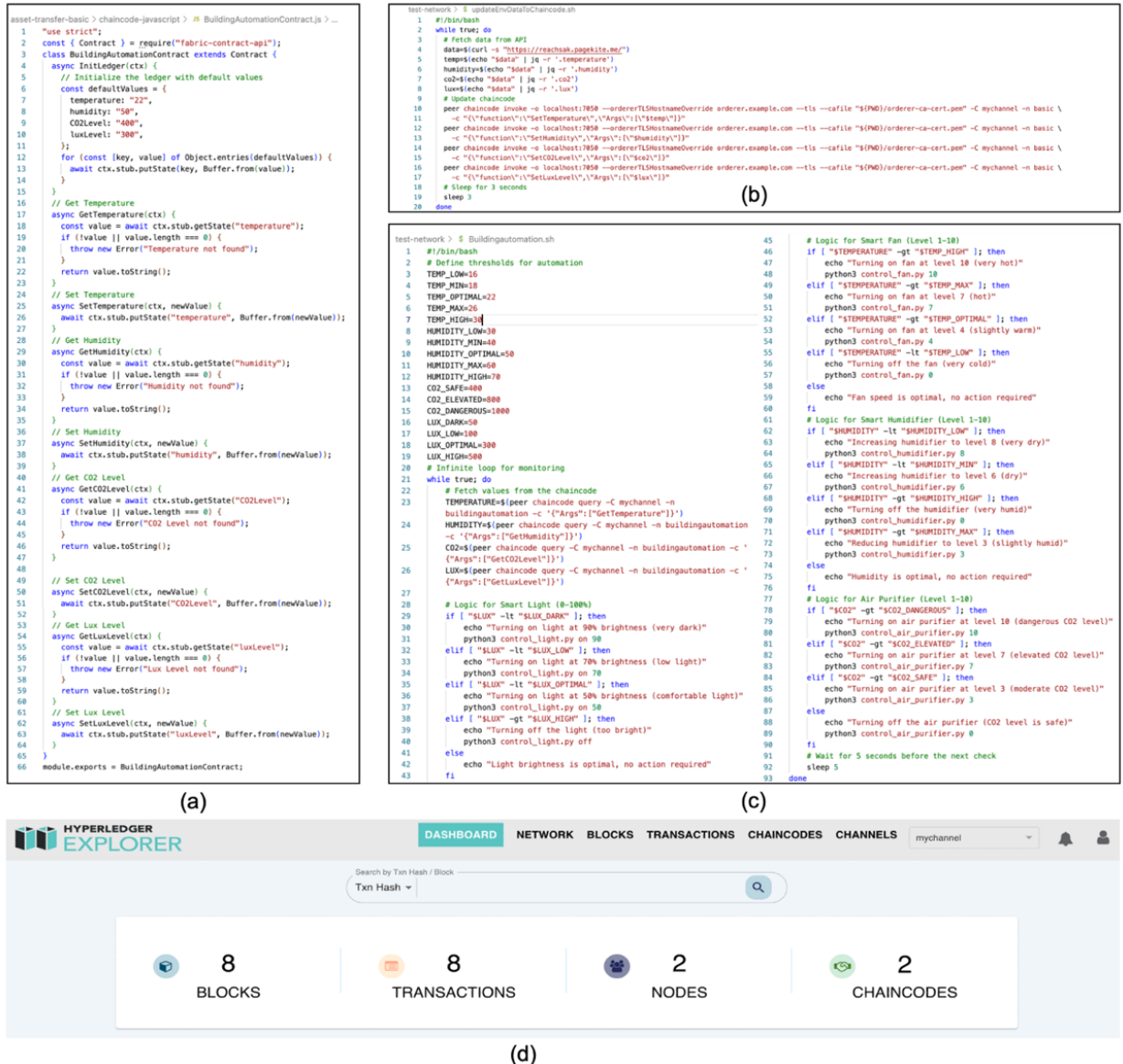

**Fig. 10** Overview of the Hyperledger Fabric-based system for decentralized automation of building operation

### 5.2.2. Hyperledger fabric-based building automation system

The Hyperledger Fabric-based building automation system integrates real-time data acquisition, chaincodes, and automated control of smart devices to optimize indoor environmental conditions. This system leverages a Raspberry Pi and sensors as the primary tools for collecting environmental sensors for temperature, humidity, carbon monoxide levels, and light intensity. The collected data is then transmitted through a REST API to the Hyperledger Fabric blockchain network for building automation applications. Fig. 10 illustrates the overall architecture of the system.

The primary components of the system include the chaincode and two bash scripts for environmental data retrieval and query into and from the blockchain as well as the logic for smart home appliance control logic (Fig. 9). The chaincode is responsible for storing and retrieving environmental parameters. It initializes the ledger with default values for temperature, humidity, carbon monoxide level, and light intensity. Each parameter is associated with Set and Get functions, allowing real-time updates and queries to the blockchain (Fig. 9 (a)). This ensures a tamper-resistant and immutable record of environmental data.

The first Bash script automates the process of fetching real-time data from Raspberry Pi's REST API and recording it into the Hyperledger Fabric blockchain. The script uses curl to retrieve sensor readings, including temperature, humidity, carbon monoxide level, and light intensity, and employs the peer chaincode invoke command to store the data in the blockchain. This process is executed in an infinite loop with a 60-second interval, ensuring consistent synchronization of real-world data with the blockchain ledger (Fig. 9 (b)). The second Bash script queries the stored data from the blockchain and implements decision-making logic to control smart devices, such as lights, fans, and humidifiers. Using peer chaincode query, it retrieves current values of temperature, humidity, carbon monoxide level, and light intensity (Fig. 9 (c)). The script defines specific thresholds for each parameter, which are used to trigger actions on the smart devices. For instance, if the light intensity is below a defined threshold, the script invokes another Python script for lighting control to adjust the brightness of the lighting system. Similarly, temperature readings influence the operation levels of smart fans, while humidity thresholds dictate the activity of smart humidifiers.

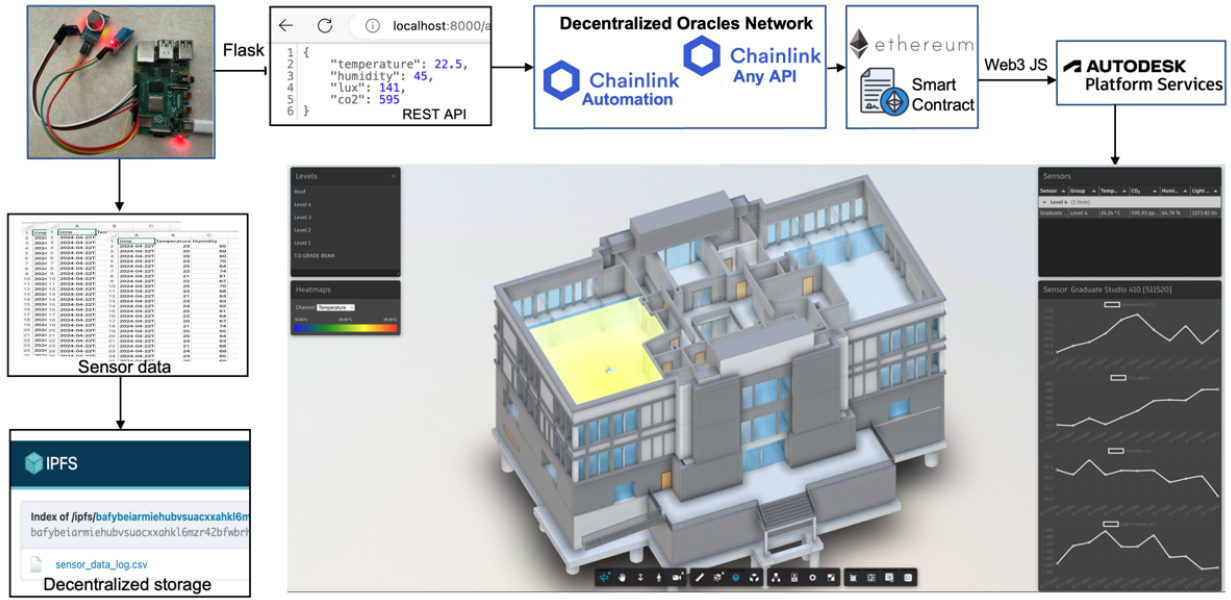

**Fig. 11** Overview of the Ethereum-based system for decentralized digital twin

### 5.3. Case study 2
#### 5.3.1. Ethereum-based digital twin

The Ethereum-based digital twin prototype leverages a combination of hardware, decentralized oracle networks (DONs), blockchain technology, and digital twin platforms to create a smart, automated, and transparent system for visualizing building environmental conditions. This system was developed using a Raspberry Pi equipped with environmental sensors, REST API integration, Chainlink's decentralized oracle network, Web3.js, and Autodesk Tandem. Fig. 11 illustrates the overview of the system.

In this prototype, the environmental data is initially captured by sensors connected to the Raspberry Pi. A Flask-based REST API is developed to serve this sensor data in JSON format, making it accessible to external applications. To bridge the gap between the off-chain sensor data and the Ethereum blockchain, the prototype utilized the decentralized oracle network. Decentralized oracle networks provide a secure and reliable mechanism to bring external data onto the blockchain, ensuring decentralization and trustworthiness in the data retrieval process.

Two of Chainlink's decentralized oracle network components are utilized within this prototype: Chainlink Any API and Chainlink Automation. The configuration of these two components is illustrated in Fig. 12 Chainlink Any API enables smart contracts to access data from any API outside of the blockchain network. In this Ethereum-based digital twin system, Chainlink Any API is used within the MultiWordConsumer contracts to request and retrieve the four environmental parameters from the REST API created by Raspberry Pi (Fig. 13 (a)). The contract includes the requestMultipleParameters function, which is designed to handle requests for multiple data points from the REST API. The contract was deployed on the Ethereum Sepolia testnet.

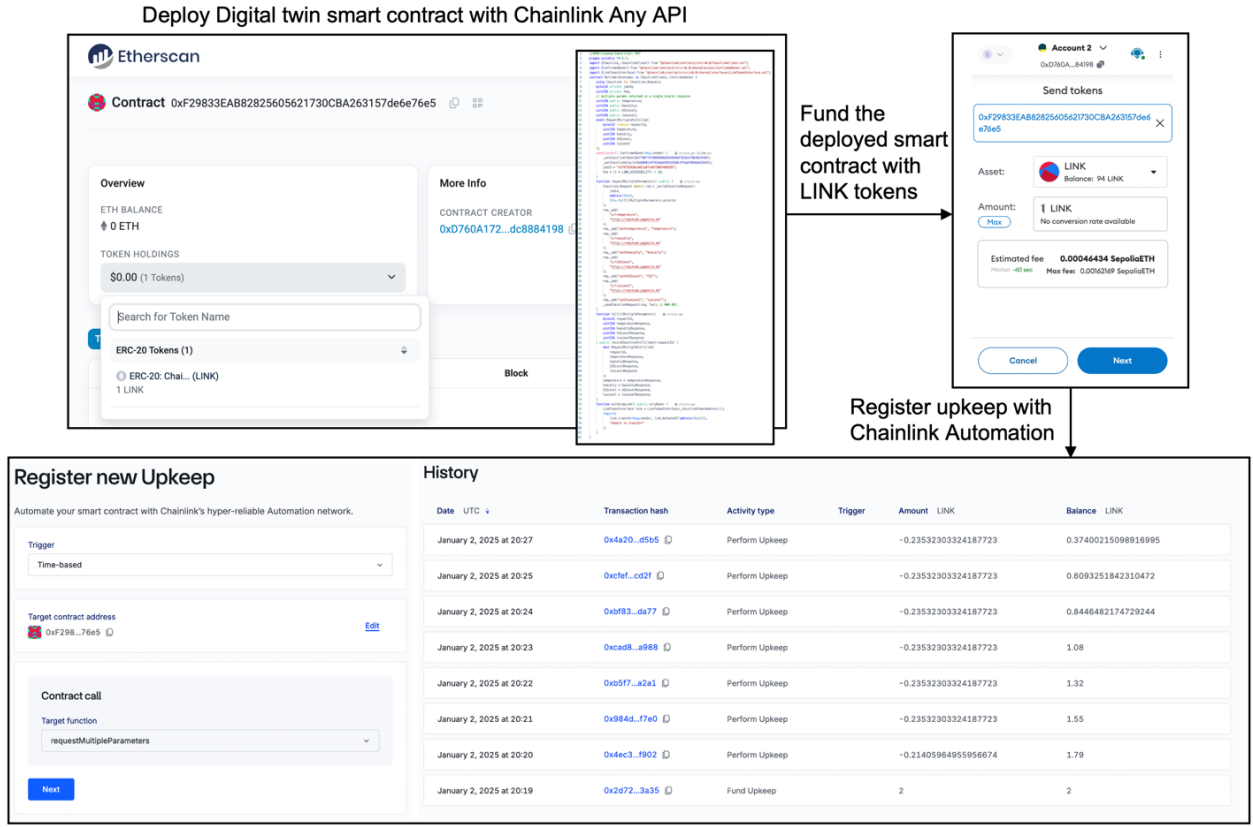

**Fig. 12** Configuration of Chainlink Any API and Chainlink Automation

For the digital twin implementation in the Ethereum-based system, Autodesk Platform Services is used instead of the prebuilt solutions like Autodesk Tandem. In this implementation, the Model Derivative API provided by Autodesk Platform Services enables translation of BIM models and transmission of real-time building parameters, such as temperature, humidity, carbon monoxide levels, and light intensity levels, to create a visualized digital twin model. The system also incorporates time-based automation using Chainlink Automation. Chainlink Automation provides the ability to automate Ethereum smart contract functions based on predefined logic or time intervals. In this case, the requestMultipleParameters function of the MultiWordConsumer contract is automatically triggered at regular 60-second intervals. Additionally, the prototype also incorporates the periodic storage of historical sensor data from the Raspberry Pi to the IPFS.

For the DON components, such as the MultiWordConsumer contract and Chainlink Automation, to function effectively, the smart contract must be funded with LINK tokens, which are the native utility tokens of the Chainlink network. These tokens are used to compensate the oracles for their services and to enable functionalities within the Chainlink ecosystem. Faucet LINK tokens were used in this experiment. Faucet tokens are free test tokens distributed on blockchain testnet to enable developers to experiment with and test their smart contracts and decentralized applications without incurring financial costs.

**Fig. 13** a) Smart contract for decentralized Digital twin with Chainlink Any API b) Smart contract for decentralized building automation system

### 5.3.2. Ethereum-based building automation system

The Ethereum-based building automation system builds upon the Ethereum-based digital twin system described in the previous section. This system integrates additional components such as smart home appliances, smart contracts for building automation thresholds, and Python script to achieve automated and decentralized control of smart building appliances. The overview of this automation system is presented in Fig. 14.

The first smart contract used in this prototype is the MultiWordConsumer contract, which was previously described in the previous section. This contract continuously updates and stores the latest environmental parameters from the Raspberry Pi's REST API. As in the previous section, Chainlink Any API is used to fetch data from the REST API, while Chainlink Automation ensures periodic updates to the contract. Chainlink Automation is also used to trigger the requestMultipleParameters function of the MultiWordConsumer contract every 60 seconds.

In addition, this system also includes the BuildingAutomationConfig smart contract (Fig. 13 (b)), which defines the baseline comfort parameters for the building environment. This contract includes eight variables representing the minimum and maximum allowable values for temperature, humidity, carbon monoxide levels, and light intensity levels. The contract also provides setter and getter functions to modify and retrieve these baseline parameters, ensuring flexibility in building automation logic.

In addition, a Python script with the Web3.py library was developed to extract data from these two smart contracts for building automation. This script retrieves real-time environmental data from the MultiWordConsumer contract and compares it with the baseline comfort parameters stored in the BuildingAutomationConfig contract. If any real-time data falls outside the defined comfort range, the script sends control signals to the corresponding smart appliances in the building for adjustment. For instance, if the real-time temperature exceeds the maximum threshold, the script can activate cooling systems to bring the temperature back within the acceptable range. Conversely, if the illuminance level drops below the minimum threshold, the script can increase the brightness of the smart lighting system. The goal is to continuously adjust the building's environmental conditions to maintain them within the predefined comfort ranges, thereby optimizing indoor comfort and energy efficiency.

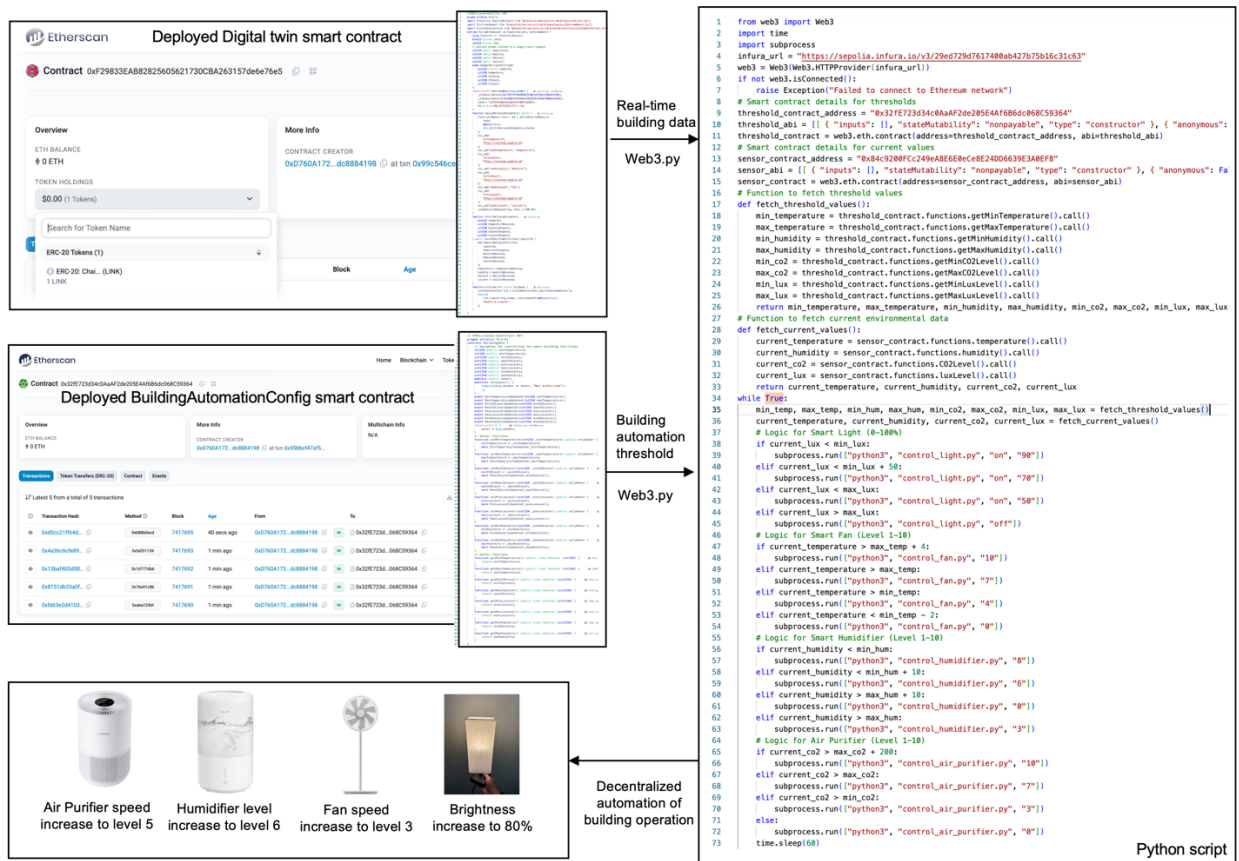

**Fig. 14** Overview of the Ethereum-based system for decentralized automation of building operation

6. Result, Evaluation, and Discussion
6.1. Cost analysis

This section presents the financial implications of the two blockchain systems presented in the case studies in the previous section. The Hyperledger Fabric-based system does not incur any operational costs for the blockchain transaction. Instead, network participants are pre-authorized, and transactions are verified with Raft or Kafka consensus mechanisms (Yang et al. 2022), which do not involve fees. In contrast, the Ethereum-based system incurs transaction costs in the form of gas fees, which serve as compensation for the computational resources required to process and validate operations on the blockchain. These include smart contract deployments and the decentralized oracle network. These fees are expressed in Ether (ETH) and converted to USD using the ETH-USD exchange rate as of January 2025. For instance, deploying the primary smart contracts, such as the BuildingAutomationConfig contract and the Digital Twin contract, consumed significant amounts of gas. These deployments collectively cost approximately 0.031848 ETH (~104.54 USD). Additionally, the Digital Twin Contract (MultiWordConsumer contract) and the Chainlink Automation service required LINK tokens. Each transaction for the decentralized data transmission from Raspberry Pi to the digital twin platform incurred a total cost of 0.335323 LINK, which amounts to 6.76 USD per transaction. These calculations were based on testing conducted on the Sepolia testnet, which simulated the Ethereum main net environment but did not incur actual monetary costs. The key operations with their associated costs, are summarized in Table 2. Ethereum's public network incurs high gas fees for transactions, making it less cost-efficient for IoT and digital twin applications with high transaction volumes.

**Table 2** The transaction cost of the Ethereum-based digital twin and building automation system

| Operations | Smart contract/Service | Gas | Transaction fee | Transaction fee (USD) |
|---|---|---|---|---|
| Contract deployment | BuildingAutomationConfig contract | 821,489 | 0.009776 ETH | 32.10 |
| Contract deployment | Digital twin contract | 1,857,505 | 0.022072 ETH | 72.44 |
| Digital twin contract fund | Digital twin contract | - | 0.1 LINK | 2.02 |
| Real-time data retrieval from REST API to Digital twin contract | Chainlink automation | - | 0.235323 LINK | 4.74 |

### 6.2. Scalability

Scalability in blockchain-based systems is primarily measured through transaction latency and throughput. Private blockchains, such as Hyperledger Fabric, generally offer higher throughput and lower transaction latency due to their smaller number of nodes. This makes private blockchains more suitable for applications with high transaction volumes, as they can process data more quickly and reliably. Conversely, public blockchain platforms, such as Ethereum, often encounter scalability challenges due to the need for global consensus among distributed nodes. This results in slower transaction speeds, higher latency, and increased operational costs, especially under high transaction loads.

In this study, the scalability of the Hyperledger Fabric-based system was benchmarked using Hyperledger Caliper (Kaushal and Kumar 2024). The results presented in Fig. 15 demonstrate that the system achieved a throughput of approximately 591 transactions per second with an average latency of 0.03 seconds when performing the GetBuildingData functions. In contrast, the Ethereum-based digital twin and building automation system operates on a public blockchain infrastructure, where scalability is constrained by the proof-of-stake consensus mechanism. Every transaction on Ethereum requires validation from all participating nodes before it is appended to the blockchain. While PoS improves energy efficiency compared to proof-of-work, it still limits the transaction throughput to around 30 transactions per second (Abrol 2022).

This constraint suggests that as the number of smart appliances or buildings increases, delays in transaction processing could arise which could hinder the system's responsiveness and efficiency. While this throughput may suffice for small-scale deployments or experimental setups, it would likely become a bottleneck in large-scale, real-world scenarios involving multiple users or a dense network of smart home appliances.

The performance analysis of the two developed systems underscores the superior scalability, and speed of the Hyperledger Fabric-based system compared to the Ethereum-based system. With its higher transaction throughput and lower latency, Hyperledger Fabric is well-suited for real-time IoT data transfer for digital twin and building automation systems.

```
+----------------+-------+------+-----------------+------------------+------------------+------------------+------------------+
| Name           | Succ  | Fail | Send Rate (TPS) | Max Latency (s)  | Min Latency (s)  | Avg Latency (s)  | Throughput (TPS) |
|----------------|-------|------|-----------------|------------------|------------------|------------------|------------------|
| GetBuildingData| 17489 | 0    | 591.0           | 0.03             | 0.00             | 0.00             | 590.9            |
+----------------+-------+------+-----------------+------------------+------------------+------------------+------------------+

2024.03.05-02:35:09.177 info  [caliper] [round-orchestrator]    Finished round 1 (GetBuildingData) in 30.101 seconds
2024.03.05-02:35:09.177 info  [caliper] [monitor.js]    Stopping all monitors
2024.03.05-02:35:09.177 info  [caliper] [report-builder]        ### All test results ###
2024.03.05-02:35:09.177 info  [caliper] [report-builder]
+----------------+-------+------+-----------------+------------------+------------------+------------------+------------------+
| Name           | Succ  | Fail | Send Rate (TPS) | Max Latency (s)  | Min Latency (s)  | Avg Latency (s)  | Throughput (TPS) |
|----------------|-------|------|-----------------|------------------|------------------|------------------|------------------|
| GetBuildingData| 17489 | 0    | 591.0           | 0.03             | 0.00             | 0.00             | 590.9            |
+----------------+-------+------+-----------------+------------------+------------------+------------------+------------------+
```

**Fig. 15** Hyperledger Caliper benchmark results

### 6.3. Security and privacy

The proposed Hyperledger Fabric-based system and Ethereum-based system offer a significant improvement over BACnet, the traditional building automation protocol, in terms of system security. BACnet's centralized architecture is inherently vulnerable to several cyberattacks, including man-in-the-middle attacks, data tampering, and denial-of-service attacks (Peacock et al. 2018; Morales-Gonzalez et al. 2024). For example, an attacker could intercept or alter communication between devices in a BACnet network, disrupt system operations through a denial-of-service attack, or compromise the central controller to manipulate critical building functions.

In contrast, both the Hyperledger Fabric-based system and the Ethereum-based system employ decentralized architectures, which inherently mitigate many of these vulnerabilities. The distributed nature of Hyperledger Fabric also eliminates the single point of failure, making it resistant to denial-of-service attacks (Jayadev et al. 2024). Its permissioned blockchain model enforces strict access controls, preventing unauthorized entities from participating in the network. Additionally, transactions within the Hyperledger Fabric network are verified through consensus mechanisms, ensuring data integrity, and preventing tampering (Honar Pajooh et al. 2021). Furthermore, cryptographic techniques employed in the Ethereum blockchain systems, such as digital signatures and hashing, provide an additional layer of security, effectively protecting IoT data from interception and manipulation (Raj et al. 2021). In addition, databases for IoT data such as InfluxDB and TimescaleDB rely on centralized servers, which could make them susceptible to cyber-attacks, data breaches, data loss, and tampering. The IPFS is used as a decentralized storage solution in this study. IPFS distributes data across multiple nodes which enhances data integrity and improves resistance against unauthorized access or data loss (Ali et al. 2017).

Regarding data privacy, the Hyperledger Fabric-based system offers greater control over access permissions and data visibility, which is suited for building automation applications. As a private blockchain, it allows network participants to define and enforce strict rules about who can view or interact with the data, enhancing both privacy and security. On the other hand, the Ethereum-based system, as a public blockchain, provides transparency and immutability through its decentralized consensus mechanism. While this fosters trust and ensures data integrity, it also means that data on the Ethereum network is exposed to public scrutiny, which may not be ideal for applications involving sensitive or proprietary information. This distinction highlights one aspect of the trade-offs between privacy and transparency when choosing a blockchain system for digital twins and building automation applications.

### 6.4. Limitation

The proposed blockchain-based frameworks present an innovative approach to decentralized automation of building operations and digital twins for building infrastructure. While offering significant advantages, these frameworks are not without limitations. This section outlines the primary constraints associated with the proposed systems and their proof-of-concept implementation. One key limitation of the Ethereum-based framework is the inherent volatility of the Ethereum cryptocurrency, which serves as the medium for transaction fees within the system. The fluctuating value of Ethereum introduces financial uncertainty, particularly concerning the costs of updating real-time data for the digital twin and building automation processes. This unpredictability can result in discrepancies between anticipated and actual expenses, potentially discouraging widespread adoption. To mitigate this issue, future research could consider using stablecoins, such as USDT or USDC (Thanh et al. 2023), which maintain stable value by being pegged to reserve assets like the U.S. Dollar. Stablecoins offer a more reliable decentralized payment option that could enhance system adoption and usability. Additionally, the current implementation relies on smart appliances rather than employing a fully integrated smart building automation system. While these devices were sufficient for meeting the research objectives and demonstrating the framework's capabilities, they highlight a gap in achieving a comprehensive, end-to-end smart building solution. Future work could address this limitation by integrating more advanced and holistic building systems into the framework. Additionally, any required updates to system logic, environmental parameters, or the source of REST APIs within the proposed framework will also require redeployment to the respective blockchain networks. This process can be time-consuming and expensive. Future research could consider using upgradable smart contracts and modular design approaches to address this limitation.

## 7. Future research

This study represents the first attempt to employ a Hyperledger Fabric and Ethereum blockchain-based system alongside the decentralized oracle network to develop a decentralized digital building twin and a decentralized building automation system. In the current implementation, the automation logic in both the Hyperledger Fabric and Ethereum-based digital twin systems can be adjusted manually by updating parameter thresholds in the smart contracts. This capability opens new research avenues into the decentralized control of building automation systems through technologies like Decentralized Autonomous Organizations (DAOs) (Wang et al. 2019). DAOs could allow stakeholders to collectively manage and govern building operations in a decentralized manner, ensuring transparency and equity in decision-making processes. While there is increasing research on DAO implementations for Ethereum, there remains limited exploration of DAO frameworks for private blockchains like Hyperledger Fabric. This gap also presents a significant opportunity to investigate how DAOs can be tailored for private blockchain environments to facilitate decentralized building automation systems (Ly et al. 2024). The potential customization of such frameworks for governance in building infrastructure or even large-scale civil infrastructure could transform how systems are democratically controlled and autonomously operated by DAO members.

The proposed framework also demonstrates the capability to store historical building data on decentralized storage systems like IPFS. This feature could support future applications in decentralized data marketplaces, where building owners might monetize their building data through web3 technologies (Ramachandran et al. 2018; Nawaz et al. 2020). For instance, environmental and operational data stored on IPFS could be sold to organizations or researchers seeking datasets for AI training or other research purposes (Tara et al. 2023).

Finally, the integration of decentralized building automation systems with generative AI-based agents offers another exciting direction for future research. These AI agents could leverage historical building data to understand, predict, and optimize environmental variables within a building. By autonomously adjusting parameters stored in smart contracts, these agents could establish a form of decentralized artificial intelligence for smart buildings. Emerging trends in smaller, more efficient language models, such as Microsoft's Phi-3, Meta's LLaMA 3, and Google's Gemini, suggest that sophisticated AI systems could soon be deployed on low-cost edge devices like Raspberry Pi which are capable of delivering robust performance at minimal cost (Ly and Shojaei 2024). This development holds immense potential for creating decentralized, AI-driven smart building systems. As AI capabilities continue to evolve, their integration with blockchain-based frameworks could significantly enhance the intelligence and autonomy of decentralized building operations.

## 8. Conclusion

This study presents a novel framework for decentralized digital building twins and building automation systems in smart buildings, utilizing both private and public blockchains. The Hyperledger Fabric-based framework integrates key components such as environmental data collection tools, chaincode for digital building twins and automation, and the Autodesk Tandem digital twin platform. Similarly, the Ethereum-based framework incorporates data collection components, smart contracts for decentralized building automation and digital twins, decentralized oracle networks such as Chainlink Any API and Chainlink Automation, and digital twin platforms such as the Autodesk platform service. Prototypes of both frameworks were tested in a real-world building environment with smart building appliances to validate practical applications. The evaluation criteria included cost efficiency, scalability, data security, and privacy. The findings indicate distinct strengths and limitations for each platform. Hyperledger Fabric demonstrates superior scalability, transaction speed, and cost efficiency, which is suitable for the transmission of large amounts of IoT data transfer and building automation. In contrast, Ethereum offers greater decentralization, making it suitable for applications prioritizing openness and resilience, albeit at the cost of higher transaction fees, lower throughput, and increased latency. Both frameworks enhance the security of IoT data transfer and building automation compared to traditional centralized systems like the BACnet protocol. The contributions of this research to the existing body of knowledge are as follows: (1) presenting a novel framework for decentralized IoT data transmission to the digital twin using private and public blockchain with decentralized oracle networks (2) Developing autonomous and decentralized automation of building operations using chaincode and smart contracts. (3) Developing public and private blockchain-based digital building twin (4) Analyzing the advantages and limitations of public and private blockchain applications for decentralized digital twin and building automation and comparing them against the traditional centralized systems. This framework provides a foundation for future research within the domain of web3 technology applications for smart buildings by offering a blueprint for implementing decentralized, secure, and efficient building automation and digital building twin systems.

**Acknowledgments**
This research has received no external funding.

**Availability of data and materials**
Data will be made available on request.

**Competing interests**
The authors have no conflicts of interest to declare that are relevant to the content of this article.

**Funding**
No funding was received to assist with the preparation of this manuscript.

**Authors' contributions**
Supervision: [Alireza Shojaei]; Conceptualization: [Reachsak Ly]; Writing - original draft preparation: [Reachsak Ly]; Writing - review and editing: [Reachsak Ly]; Visualization: [Reachsak Ly]; Methodology: [Alireza Shojaei]; Project administration: [Alireza Shojaei]